\documentclass[galaxies,article,submit,moreauthors,pdflatex,10pt,a4paper]{mdpi} 
\graphicspath{{Definitions/}}
\usepackage{xcolor}
\usepackage{mathalfa}
\usepackage{hyperref}
\usepackage{amsmath,amssymb}
\usepackage{multirow}
\DeclareFontFamily{U}{rcjhbltx}{}
\DeclareFontShape{U}{rcjhbltx}{m}{n}{<->rcjhbltx}{}
\DeclareSymbolFont{hebrewletters}{U}{rcjhbltx}{m}{n}
\let\aleph\relax\let\beth\relax
\let\gimel\relax\let\daleth\relax
\DeclareMathSymbol{\aleph}{\mathord}{hebrewletters}{39}
\DeclareMathSymbol{\beth}{\mathord}{hebrewletters}{98}
\DeclareMathSymbol{\gimel}{\mathord}{hebrewletters}{103}
\DeclareMathSymbol{\daleth}{\mathord}{hebrewletters}{100}
\DeclareMathSymbol{\lamed}{\mathord}{hebrewletters}{108}
\DeclareMathSymbol{\mem}{\mathord}{hebrewletters}{109}
\DeclareMathSymbol{\ayin}{\mathord}{hebrewletters}{96}
\DeclareMathSymbol{\tsadi}{\mathord}{hebrewletters}{118}
\DeclareMathSymbol{\qof}{\mathord}{hebrewletters}{113}
\DeclareMathSymbol{\shin}{\mathord}{hebrewletters}{152}
\preto{\abstractkeywords}{\nolinenumbers}

\def\mbh{$M_{\rm BH}$\/}

\def\lledd{$L/L_{\rm Edd}$}

\def\feiiq{\rm Fe{\sc ii}$\lambda$4570\/}
\def\msol{M$_\odot$\/}

\def\ltsima{$\; \buildrel < \over \sim \;$}
\def\ltsim{\lower.5ex\hbox{\ltsima}}  
\def\gtsima{$\; \buildrel > \over \sim \;$}

\def\gtsim{\lower.5ex\hbox{\gtsima}}

\def\civ{{\sc{Civ}}\/}

\def\cm3{cm$^{-3}$\/}
\def\hb{{\sc{H}}$\beta$\/}

\def\hbbc{{\sc{H}}$\beta_{\rm BC}$\/}
\def\hbsbc{{\sc{H}}$\beta_{\rm SBC}$\/}
\def\hbvbc{{\sc{H}}$\beta_{\rm VBC}$\/}
\def\hbblue{{\sc{H}}$\beta_{\rm BLUE}$\/}
\def\hbnc{{\sc{H}}$\beta_{\rm NC}$\/}
\def\mgii{{Mg\sc{ii}}$\lambda$2800\/}

\def\oiiiopt{{\sc{[Oiii]}}$\lambda\lambda$4959,5007\/}
\def\oii{{\sc{[Oii]}}$\lambda$3727\/}

\def\heiiuv{He{\sc{ii}}$\lambda$1640}

\def\feii{{Fe\sc{ii}}\/}

\def\fe{{\sc{Fe}}\/}

\def\fe76087{{\sc [Fe vii]}$\lambda$6087\/}
\def\oiii{{\sc [Oiii]}$\lambda$5007}
\def\oiiinc{{\sc [Oiii]}$\lambda 5007_\mathrm{NC}$\/}
\def\oiiisbc{{\sc [Oiii]}$\lambda 5007_\mathrm{SBC}$\/}

\def\kms{km~s$^{-1}$}

\def\ergss{erg\, s$^{-1}$\/}

\usepackage{enumerate}
\definecolor{darkorange}{rgb}{1,0.612,0}
\definecolor{aquamarine}{rgb}{0.498,1,0.8314}

 \defcitealias{sulenticetal02}{S02}
  \defcitealias{marzianietal09}{M09}
  \defcitealias{marzianietal13}{M13}

\Title{Isolating an outflow component in single-epoch spectra of quasars}
\Author{{Paola Marziani}$^{1}$\orcidA{}, Alice Deconto Machado$^{2}$\orcidB{}, Ascension Del Olmo$^{2}$\orcidC{} }
\AuthorNames{Paola Marziani}
\address{$^{1}$ \quad National Institute for Astrophysics (INAF), Astronomical Observatory of Padova, IT-35122 Padova, Italy\\
$^{2}$ \quad Instituto de Astrofis\'{\i}ca de Andaluc\'{\i}a, IAA-CSIC, {Glorieta}  de la Astronomia s/n, E-18008 Granada, Spain, chony@iaa.es (A.d.O.)\\  
}
\corres{Correspondence: paola.marziani@inaf.it; Tel.: +39-0498293415}
\abstract{Gaseous outflows appear to be a  universal property of type-1 and type-2 active galactic nuclei (AGN). The main diagnostic is provided by emission features shifted to higher frequency via the Doppler effect, implying that the emitting gas is moving toward the observer.  However, beyond the presence of blueshift, the observational signatures of the outflows are  often unclear, and no established criteria exist to isolate the outflow contribution in the integrated, single-epoch spectra of type-1 AGN. The emission spectrum collected with the typical apertures of long-slit spectroscopy or of fiber optics sample contributions over a broad range  of spatial scales, making it difficult to analyze the line profiles in terms of different kinematic components. Nevertheless, hundred of thousands of quasars spectra collected at moderate resolution demand a proper analysis of the line profiles for proper dynamical modelling of the emitting regions. In this small contribution we shall analyze several  profiles  of the {\sc Hi} Balmer line \hb\ from composite and individual optical spectra of sources radiating at moderate Eddington ratio (Population B following \citeauthor{sulenticetal00a} \citeyear{sulenticetal00a}). Features and profile shapes that might be traced to outflow due to  narrow-line region gas are detected over a wide range of luminosity. } 
\keyword{active galactic nuclei; optical  spectroscopy; ionized~gas; broad line region}
\begin{document}
%%%%%%%%%%%%%%%%%%%%%%%%%%%%%%%%%%%%%%%%%%
%% Only for the journal Gels: Please place the Experimental Section after the Conclusions

\section{Introduction}
%\vspace{-6pt}
%\subsection{Quasar Spectra: still open to interpretation}

Type-1 active galactic nuclei (AGN) are characterized by the presence of broad and narrow optical and UV lines (for introductions see e.g., \cite{osterbrockmathews86,netzer90,peterson97,sulenticetal00a,osterbrockferland06,marzianietal06}). Spectra  show a mind-boggling variety of broad emission line profiles not only among different objects, but also among different lines in the spectrum of the same object. \citet{sulentic89} carried out measurements  of spectral shifts and asymmetries exhibited by the broad lines relative to the narrow ones, proposing an empirical classification scheme for the broad {\sc Hi} Balmer line \hb. Among the classes identified by \citet{sulentic89}, two stand out: AR,R and AR,B, where AR means red-ward asymmetric, and the letter after the comma indicates either a shift of the line peak toward the red or the blue.  

Fast forward more than 30 years, type-1 quasars are now being contextualized on the basis of the main sequence (MS) trends \citep[e.g.,][]{sulenticetal00a,shenho14,pandaetal18}. Type-1 AGN have been grouped into two main populations, Population A and B, defined on the basis of the   Balmer line widths (more specifically of   \hb: FWHM \hb\ $\lesssim$ 4000 \kms\  for Population A; FWHM \hb\ $\gtrsim\ 4000$ \kms\  for Population B \citep{sulenticetal00a,sulenticetal11}  at low and moderate luminosity $\log L \lesssim 46$ [\ergss]). The classification of the quasar population along the MS has its main physical foundation on systematic differences in Eddington ratio \citep{marzianietal03b}: Population A sources typically have \lledd\ $\gtrsim 0.2 $, with extreme Population A sources reaching \lledd $\gtrsim 1$  \citep{marzianisulentic14}, values close  to the expected theoretical limit for super-Eddington accretion rate \citep{abramowiczetal88,mineshigeetal00,sadowski11}. Usually Pop. B sources present  lower values of Eddington ratio when compared with the ones of Pop. A. The governing parameter of the MS itself appears to be Eddington ratio convolved with the effect of orientation \citep[e.g.,][]{sunshen15,pandaetal19}.\footnote{In flux limited samples Pop. A and B may have similar luminosity distributions. If this is the case Pop. B sources are expected to host more massive black holes, considering the systematic differences in Eddington ratio. } 

Sources showing prominent \hb\ red asymmetries (i.e., AR,R according to \citet{sulentic89}) are classified as belonging to Population B  \citep{marzianietal03b,sulenticetal11}. The red asymmetry itself can be considered as a defining feature of Population B sources, hinting at the presence of a ``very broad component’’ (VBC) at the line base \citep{petersonferland86,marzianisulentic93,sulenticetal00c,wangli11,punsly13,wolfetal20}. The physical properties of the region associated with the VBC are largely undetermined \citep[e.g.,][]{sneddengaskell07} but the general consensus is that the region is located at the innermost radii of the broad line region (BLR), closest to the central continuum source. This inference follows from the deduction of a velocity field dominated by virial motions, at least for several population B sources \citep{petersonwandel99,petersonetal04}. The dynamical conditions of the ``very broad line region’’ (VBLR) are subject of current debate \citep{punslyetal20}. Two main alternatives have been proposed: infall and obscuration \citep{wangetal17}, or gravitational and transverse redshift \citep{gaskell88,corbin95,popovicetal95,gavrilovicetal07,bonetal15,punslyetal20}. Both mechanisms are however still consistent with a virial velocity field as the main broadening factor. 

Gaseous outflows appear to be ubiquitous in type-1 AGN, although their traceability and their kinetic power varies greatly along the main sequence \citep{marzianietal12,marzianietal16}. The signature of outflows in the optical and UV spectra is provided by the blueshift of emission lines with respect to the rest frame, under the assumption that the shift is due to Doppler effect on the wavelength of lines emitted by gas moving toward us, and that the receding side of the flow is mainly hidden from view \citep[e.g., ][]{leighlymoore04}. While there is unambiguous evidence of outflows from the emitting regions of quasars radiating at high Eddington ratio, the situation is by far less clear for Pop. B where the accretion rate is modest, as implied by the Eddington ratio $\lesssim 0.2$. High-resolution X-ray and ultraviolet (UV) observations of the prototypical Population B source NGC 5548 reveal a persistent ionized outflow traced by UV and X-ray absorption and emission lines \citep{kaastraetal14}. However, the \civ\ emission line profile lacks strong evidence of such an outflow, also because of the prominent red line wing merging with \heiiuv\ \citep{fineetal10}.  

In this short note, we address the very specific issue of the origin of sources showing a blueshift at the peak of the \hb\ emission line i.e., of the AR,B classification. The focus is on the \hb\ line because the line is a singlet, and its peak is isolated from other contaminants, offering a clear view of its broad and narrow components. The \oiiiopt\ lines recorded along with \hb\  help to assess the nature of the \hb\ line profile. In addition, the narrow, high-ionization \oiii{} emission lines are known to be affected by outflows, as indicated by the frequent blueward asymmetries and even systematic shifts \citep{whittle85,bennertetal02,komossaetal08,zamanovetal02,marzianietal16a}, Section \ref{data} presents the data used in this work, a set of composite spectra covering a wide range in luminosity and redshift, for which the \hb\ and the \oiiiopt\ emission has been covered with optical and IR spectroscopic observations. Details on how the spectral analysis was performed are shown in Section \ref{fits}. The main results come from the profile comparison of the \hb\ and \oiii\ (Section \ref{results}), and are briefly analyzed in terms of the physical conditions of the line emitting gas, as well as of the dynamical parameters of the outflow (Section \ref{discussion}).

\begin{table}
\begin{center}
\caption{Physical parameters \label{tab:phys}}\scriptsize\tabcolsep=5pt
  \begin{tabular}{lcccc} \hline
Spectrum   &  $z$  & $\log L$  & $\log$ \mbh$^{a}$   & $\log$ \lledd \\
& & [erg s$^{-1}$] &[\msol] & \\ 
\hline
\\ \hline
\multicolumn{5}{c}{Composite spectra}\\
\hline  
B1S02     &    0 -- 0.7       &    45.63$^{b}$	&	8.52	&	-1.07	\\
B1M13     &    0.4 -- 0.7 &	46.31$^{b}$	&	9.19	&	-1.06	\\
B1M09     &    0.9 -- 2.6 &	47.29$^{c}$	&	9.63	&	-0.51	\\
  \hline
\multicolumn{5}{c}{Individual, high-$L$\ quasars}\\ \hline
 
HE0001–2340    &     2.2651 &	47.09$^{c}$	&	9.78	&	-0.86	\\
Q0029+079     &    3.2798 &	47.43$^{c}$	&	9.95	&	-0.70	\\
  \hline
\multicolumn{5}{c}{Composite spectra, jetted}\\ \hline
 
B1M13CD   &      0.4 -- 0.7 &	46.51$^{b}$	&	9.39	&	-1.05	\\
B1M13FRII &        0.4 -- 0.7   &	46.62$^{b}$	&	9.44	&	-1.00	\\ \hline
  \end{tabular}
  \end{center}
\footnotesize{$^{a}$: Black hole mass computed from the \hb\ scaling law provided by \citet{vestergaardpeterson06}, using the \hb\ full profile FWHM. Applying the average correction suggested for spectral type B1 would lower the mass by a factor 0.64, and increase the \lledd\ ratio by the same factor. $^b$ Bolometric correction assumed a factor 10; $^{c}$\ Bolometric correction assumed a factor 4, as appropriate for very high luminosity sources following  \citet{netzer20}.  }  
\end{table}

\section{Data} 
\label{data}
The data analyzed in this paper refer to the most widely populated spectral type of Population B, B1, defined by  FWHM \hb\ in the range 4000 -- 8000 \kms \citep{sulenticetal02}.
Median composite spectra covering the \hb\ range were computed over the spectral type B1 sources belonging to two  samples  of low-to-moderate redshift and luminosity,  \citep[][hereafter \citetalias{sulenticetal02} and \citetalias{marzianietal13}]{sulenticetal02,marzianietal13a}, and one sample of intermediate $z$\ and high luminosity \citep[][hereafter \citetalias{marzianietal09}]{marzianietal09}. 
The \citetalias{sulenticetal02} composites are based on the individual observations of \citet{marzianietal03a} that involved 97 B1 spectra.  The \citetalias{marzianietal13} composites are SDSS spectra in the redshift range 0.4 -- 0.7, covering both \mgii\ and \hb. The radio-quiet B1 composite was computed over 179 spectra, while the CD and FR-II composites involved 16 and 23 spectra, respectively. The B1 composite of \citetalias{marzianietal09} included 22 high-luminosity, Hamburg ESO (HE) quasars.   
Median composites where constructed from continuum-normalized (at 5100 \AA) spectra, after a   determination of the heliocentric redshift based on \oii\ or narrow component of \hb, two low-ionization narrow line that provide the best estimators of the systemic redshift of the host galaxy \citep{bonetal20}. The accurate redshift correction allowed for the preservation of the spectral resolution of the individual spectra. The \citetalias{marzianietal13} composites should therefore have a resolving power $\lambda/\delta \lambda \sim 2000$. The resolving power is only slightly lower for   \citetalias{sulenticetal02}, $\lambda/\delta \lambda \sim 1000$. The HE ISAAC near-IR observations were all collected with a narrow slit (0.6 arcsec) that yielded $\lambda/\delta \lambda \sim 1000 $, comparable to the spectra of the samples observed with optical spectrometers. 
Their main properties are summarized in Table \ref{tab:phys}, where the first column lists an identification code, and the following columns list the redshift range, and median values of bolometric luminosity, black hole mass \mbh, and Eddington ratio \lledd. In addition to the composite spectra, the spectra of two quasars of extreme luminosity at intermediate redshift  (Deconto-Machado et al. 2022, in preparation) provide examples of two opposite cases, one where a prominent outflow signature is detected (Q0029+079), and one in which there is no obvious evidence of outflow (HE0001-2340). The last two lines of Table \ref{tab:phys} consider composites for core dominated (CD) and Fanaroff-Riley (FR) sources belonging to spectral type B1 from the \citetalias{marzianietal13} sample. These two composite were defined to address the somewhat controversial issue of the mild-ionized outflow presence among radio-loud, jetted AGN. \footnote{We consider the attribute ``radio-loud'' as 
synonym  of relativistically jetted \citep{padovani16,padovani17}.  } The data of Table \ref{tab:phys}   confirm that the empirical selection of spectral type B1 corresponds to the selection of modest \lledd\ radiators. At the higher redshift and luminosity, the \lledd\ appears somewhat higher (\lledd $\approx 0.3$) because of the preferential selection of higher \lledd\ for a fixed black hole mass in flux limited surveys \citep{sulenticetal14}.

\section{Analysis}
\label{fits}

The non-linear multicomponent fits were performed using the \textsc{specfit} routine from \textsc{IRAF} \citep{kriss94}. This routine allows for simultaneous minimum-$\chi^2$ fit of the continuum approximated by a power-law and the spectral line components yielding FWHM, peak wavelength, and intensity of all line components. In the optical range we fit the H$\beta$\ profile as well as the\oiiiopt\ emission lines and the  \feii\ multiplets accounted for by a scaled and broadened template \citep{borosongreen92}. The details of the multi-component analysis has been given in several previous papers \citep[e.g., ][]{sulenticetal17} and will not be repeated here. Suffice to say that the broad profiles of Pop. B sources can be successfully modelled with two Gaussians: (1) one narrower, unshifted or slightly shifted to the red; and (2) one broader, with FWHM  $\sim $ 10000 \kms, and shifted by few thousands \kms\ to the red \citep{marzianietal03c}. This model accounts for the AR,R profile type. In addition to the model decomposition, we measured several parameters on the full broad profile \citep{zamfiretal10}. The definitions of the centroids and of the asymmetry index $A.I.$ are reported below for convenience:  
   
 \begin{equation}
c(\frac{i}{4}) = \frac{v_\mathrm{r,B}(\frac{i}{4})+v_\mathrm{r,R}(\frac{i}{4})}{2} , \, i=1,2,3; \frac{i}{4} = 0.9 
,\end{equation}
where $c$ is the speed of light and the radial velocities are measured with respect to the rest frame at fractional intensities $\frac{i}{4}$\ for each value of the index $i$ on the blue and red side of the line with respect to the rest frame. 

\begin{equation}
A.I.(\frac{1}{4}) = \frac{v_\mathrm{r,B}(\frac{1}{4})+v_\mathrm{R}(\frac{1}{4})- 2v_\mathrm{r,P}}{v_\mathrm{r,R}(\frac{1}{4})-v_\mathrm{r,B}(\frac{1}{4})}.
\end{equation}  
Note that the $A.I.$, unlike the centroids, is defined as a shift with respect to the line peak radial velocity   $v_\mathrm{r,P}$\ ($v_\mathrm{r,P}$\ is measured with respect to rest frame; a suitable proxy is provided by $c(0.9)$).

%It is a 6-dimensional parameter space that is customarily not explored in full in  photoionization computations. Here we will consider trends in  $n_\mathrm{H}$,  $U$, $Z$, $N_\mathrm{c}$, assuming 0 micro turbulence and a typical AGN SED as parameterized by \cite{mathewsferland87}. The SED is appropriate for  luminous type-1 AGN, radiating at moderate Eddington ratio \citep{ferlandetal20}.  

%The code  {\tt CLOUDY} is designed to model environments that range from the low-density limit to thermodynamic equilibrium \cite{ferlandetal13,ferlandetal17}.  {\tt CLOUDY} models the ionization, chemical, and thermal state of gas exposed to a radiation field, and predicts its emission  spectra and physical parameters. In~{\tt CLOUDY}, collisional excitation and radiative processes typical of mildly ionized gases are included, at the expense of the radiation transfer, which is treated via  a mean escape probability formalism.  Maps are built on an array of 667 of  {\tt CLOUDY} 13.05  photoionization models for a given metallicity $Z$\ and $N_\mathrm{c}$, constant $n$\ and $U$\ evaluated at steps of 0.25 dex covering the ranges $7 \le \log n_\mathrm{H} \le 14$ [cm$^{-3}$], $-4.5 \le \log U \le 0$. They were repeated for 12 values of the metallicity, in the range 0.01 $Z_\odot$ -- 100 $Z_\odot$. Three values of $N_\mathrm{c}$ were considered $\log N_\mathrm{c} = 21,22,23$\  [cm$^{-2}$]. 

\section{Results}
\label{results}

\begin{figure}[t!]
\centering
\includegraphics[width=6.5 cm]{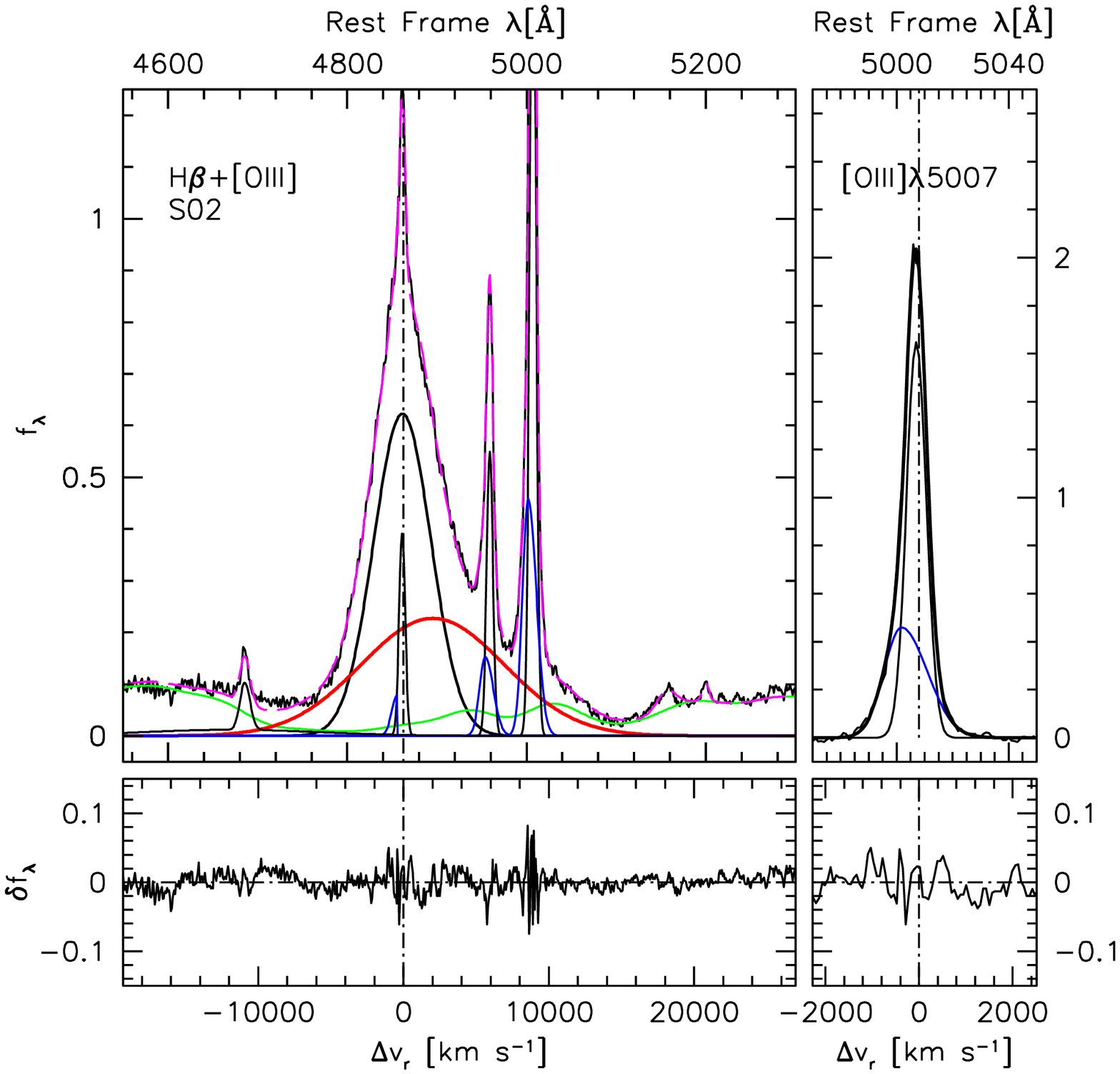}\\
\includegraphics[width=6.5 cm]{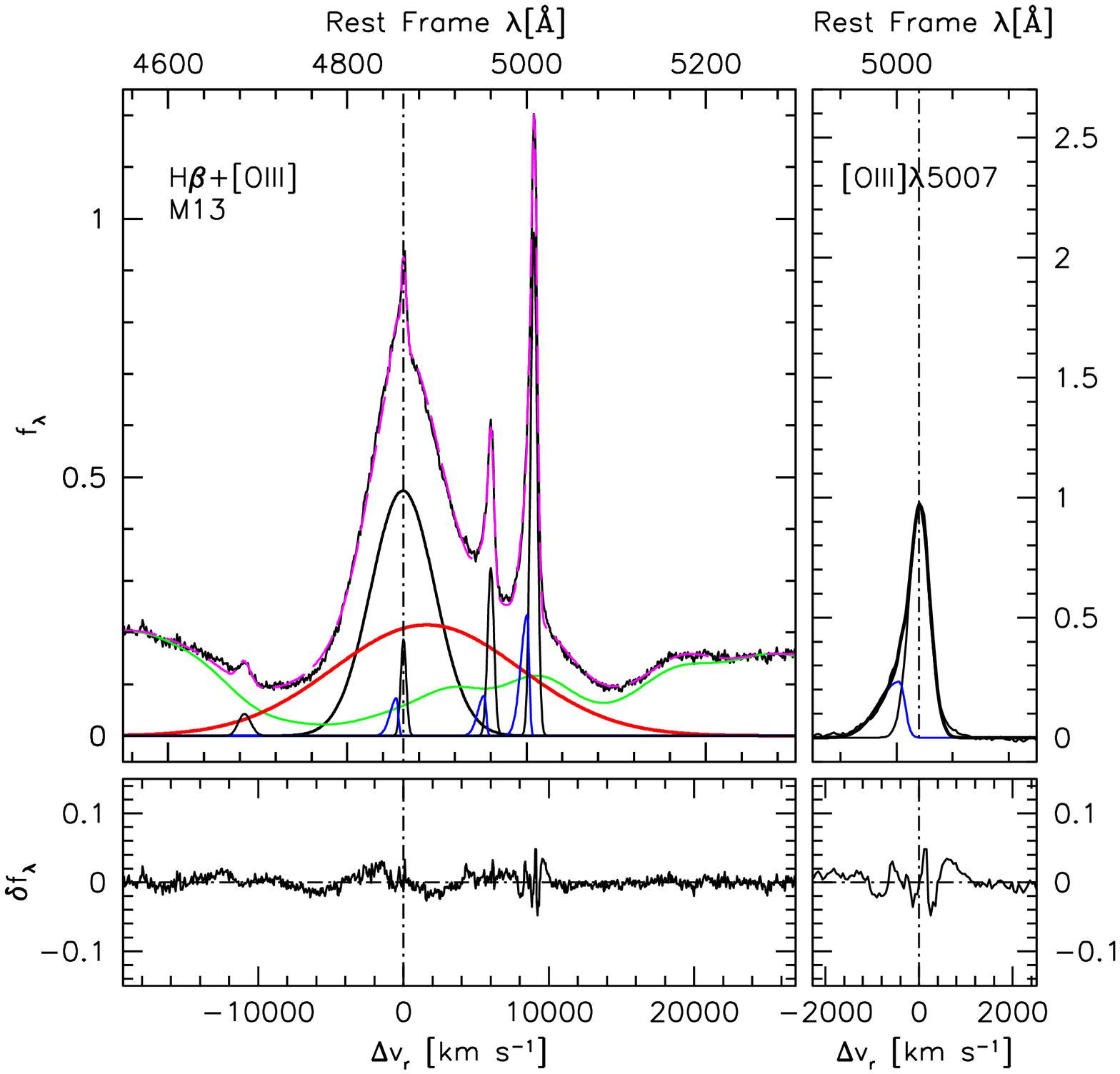}\\
\includegraphics[width=6.5 cm]{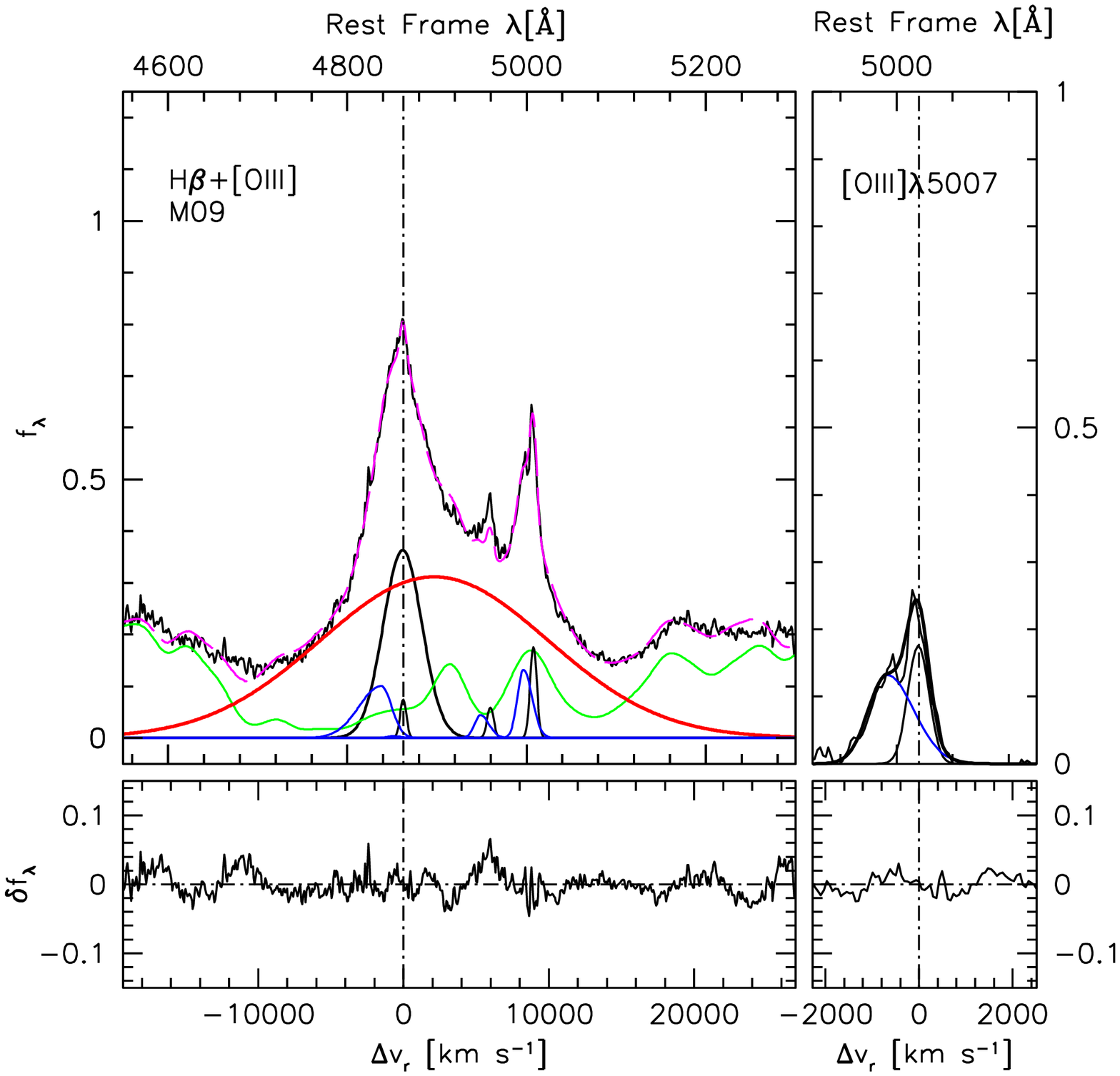}
 \caption{Analysis of the \hb\ + \oiiiopt\ region for the \citetalias{sulenticetal02} (top), \citetalias{marzianietal13} (middle) and \citetalias{marzianietal09} (bottom) B1 composite spectra. Continuum subtracted spectra are shown in the rest frame, over the range 4550 — 5300 \AA\ (left panel), with an expansion around \oiii\ (right panels). Thin solid lines: continuum-subtracted spectrum; dashed magenta line: model spectrum; thick black line: \hb\ broad component; red thick line, \hb\ very broad component;  thin smooth black lines: narrow components of \hb\ and \oiii; blue lines: blue shifted components. Green lines trace the scaled and broadened \feii\ emission template. The lower panel show the observed minus model residuals in radial velocity scale.}
\label{fig:medians} 
\end{figure}     

\paragraph{Broad \hb\ } Fig. \ref{fig:medians} shows the continuum-subtracted spectra and their models for the  three composite spectra or \citetalias{sulenticetal02}, \citetalias{marzianietal13}, and \citetalias{marzianietal09} (top, middle and bottom panel, respectively). The measurements of the broad \hb\ line  parameters are reported in Table \ref{tab:meab}. For each spectrum, Table \ref{tab:meab} lists the normalized flux $F$\ of the \hb\ full broad profile (\hbbc\ + \hbvbc + \hbblue), its equivalent width $W$ \hb\ in \AA, and the normalized fluxes of the \hbbc\ and \hbvbc\  separately.  The following columns report several parameters for the \hb\ blue shifted excess with respect to the standard Population B decomposition involving only \hbbc\ and \hbvbc: normalized flux, equivalent width, peak shift, FWHM and skew.  The last columns yield the normalized flux and the equivalent width of the \feiiq\ emission blend as defined by \citet{borosongreen92}. The equivalent width values correspond roughly to the normalized flux, so that they are reported only for the main features. The normalized fluxes can be approximately converted into luminosities by multiplying them by  the luminosity values reported in Table \ref{tab:phys} divided by the bolometric correction and by 5100 i.e., by the wavelength in \AA\ at which the continuum was normalized. Table \ref{tab:hb} reports the FWHM, A.I., and centroids as defined in \S \ref{fits} for the {\em broad} \hb\ profile (\hbbc\ + \hbvbc+ \hbblue\ i.e., without considering the narrow [\hbnc] and semi-broad [\hbsbc] components associated with narrow-line region emission). Only at the highest $L$ blueshifted emission with broad profile (\hbblue) is detected in the \hb\ profile: in this case, the \hbblue\ contribution is $\lesssim$ 5\%\ of the total line luminosity for the \citetalias{marzianietal09} and reaches  about 1/3 of the total line luminosity in the admittedly extreme Q0029 case.  In no case, however, the \hbblue\ is able to create a significant shift to the blue close to the line base: the red asymmetry dominates, and  even   the Q0029 \hb\ broad profile is ``symmeterized’’ toward the line base, with centroid at   $\frac{1}{4}$\ peak intensity close to 0 \kms.

 \begin{table}
\begin{center}
\caption{Broad-line properties measurements  \label{tab:meab}}\scriptsize\tabcolsep=2pt
 \hspace{-1cm}  \begin{tabular}{lcccccccccccccc} \hline
Spectrum	&	\multicolumn{2}{c}{\hb}	&	   \hbbc	&	\hbvbc 	&  	 	\multicolumn{4}{c}{\hbblue} &	\multicolumn{2}{c}{\feiiq	}	  &  	\\ \cline{2-3}\cline{6-9}\cline{10-11}
 		&            F       &          W$^a$	         & 	      F         &            F        &              F & Shift$^b$ & FWHM$^b$  & Skew $^c$ &       F                &                    W                     &    \\ 
%&               & [erg s$^{-1}$] &[AA] & \\ 
\hline
\\ \hline
\multicolumn{10}{c}{Composite spectra}\\
\hline  
B1S02	&	95.3		&	86.7		&	49.8	&	45.4		&	\ldots 	&	\ldots	&	\ldots	&	\ldots	&	18.1	&	14.6	 	\\
B1M13	&	122.3	&	126.5	&	52.5	&	69.8		&	\ldots	&	\ldots	&	\ldots	&	\ldots	&	47.8	&	43.0	 	\\
B1M09	&	123.3	&	129.1	&	19.6	&	99.1		&	4.6	         &	-1535	&	3611	        &	0.5		&	39.2	&	34.1	 	\\  \hline
\multicolumn{10}{c}{Individual, high-$L$\ quasars}\\ 
\hline														  					
HE0001	&	99.3	&	95.1	&	26.7	&	72.7		&	\ldots	&	\ldots	&	\ldots	&	\ldots	&	20.6	&	16.3		 	\\
Q0029	&	69.8	&	66.4	&	13.9	&	32.2		&	   23.7	&	-2097	&	4711	&	1.2			&	25.8	&	21.4                  	\\  \hline
\multicolumn{10}{c}{Composite spectra, jetted}\\ \hline
B1M13CD		&	113.8	&	118.5	&	42.9	&	70.9	&		\ldots	&	\ldots	&	\ldots	&	\ldots	&	35.6		&	32.7	 	 	\\
B1M13FRII	&	129.8	&	131.1	&	57.5	&	72.3		&	\ldots	&	\ldots	&	\ldots	&	\ldots	&	24.6		&	21.9		 	\\ \hline
\end{tabular}
 \end{center}
\footnotesize{$^a$: in units of \AA; $^b$: in units of \kms. $^{c}$ skew as reported by the  \textsc{specfit} routine; it is equal to the conventional definition of the skew \citep{azzaliniregoli12} + 1.}  
\end{table} 
\hoffset=0cm

\begin{table}
\begin{center}
\caption{\hb\ profile properties measurements \label{tab:hb}}\scriptsize\tabcolsep=5pt
 \hspace{-1cm}  
 \begin{tabular}{lcccccc} \hline
Spectrum	&	FWHM$^a$		&	AI	&	c(1/4)$^a$	&	c(1/2)$^a$	&	c(3/4)$^a$		&	c(0.9)$^a$	 	\\
\hline
\multicolumn{7}{c}{Composite spectra}\\
\hline  
B1S02	&	5560	$\pm$	170	&	0.12	$\pm$	0.03	&	680	$\pm$	230	&	250	$\pm$	80	&	160	$\pm$	70	&	130	$\pm$	50	\\
B1M13	&	6540	$\pm$	210	&	0.12	$\pm$	0.06	&	740	$\pm$	340	&	150	$\pm$	110	&	50	$\pm$	90	&	40	$\pm$	60	\\
B1M09	&	6010	$\pm$	450	&	0.28	$\pm$	0.06	&	2120	$\pm$	490	&	-50	$\pm$	220	&	-230	$\pm$	70	&	-270	$\pm$	50	\\
  \hline
\multicolumn{7}{c}{Individual, high-$L$\ quasars}\\ 
\hline	
HE0001	&	6510	$\pm$	690	&	0.29	$\pm$	0.09	&	2700	$\pm$	560	&	1310	$\pm$	340	&	900	$\pm$	170	&	830	$\pm$	110	\\
Q0029	&	6200	$\pm$	380	&	0.18	$\pm$	0.10	&	430	$\pm$	500	&	-380	$\pm$	190	&	-500	$\pm$	160	&	-500	$\pm$	110	\\
	  \hline
\multicolumn{7}{c}{Composite spectra, jetted}\\ \hline
B1M13CD	&	6880	$\pm$	240	&	0.23	$\pm$	0.06	&	1520	$\pm$	380	&	270	$\pm$	120	&	70	$\pm$	90	&	20	$\pm$	60	\\
B1M13FRII	&	6790	$\pm$	220	&	0.10	$\pm$	0.06	&	820	$\pm$	330	&	320	$\pm$	110	&	240	$\pm$	90	&	230	$\pm$	60	\\	 
\hline
\end{tabular}
  \end{center}
\footnotesize{$^a$ In units of \kms.}  
\end{table} 
\hoffset=0cm

\begin{figure}[t!]
\centering
\includegraphics[width=6.5 cm]{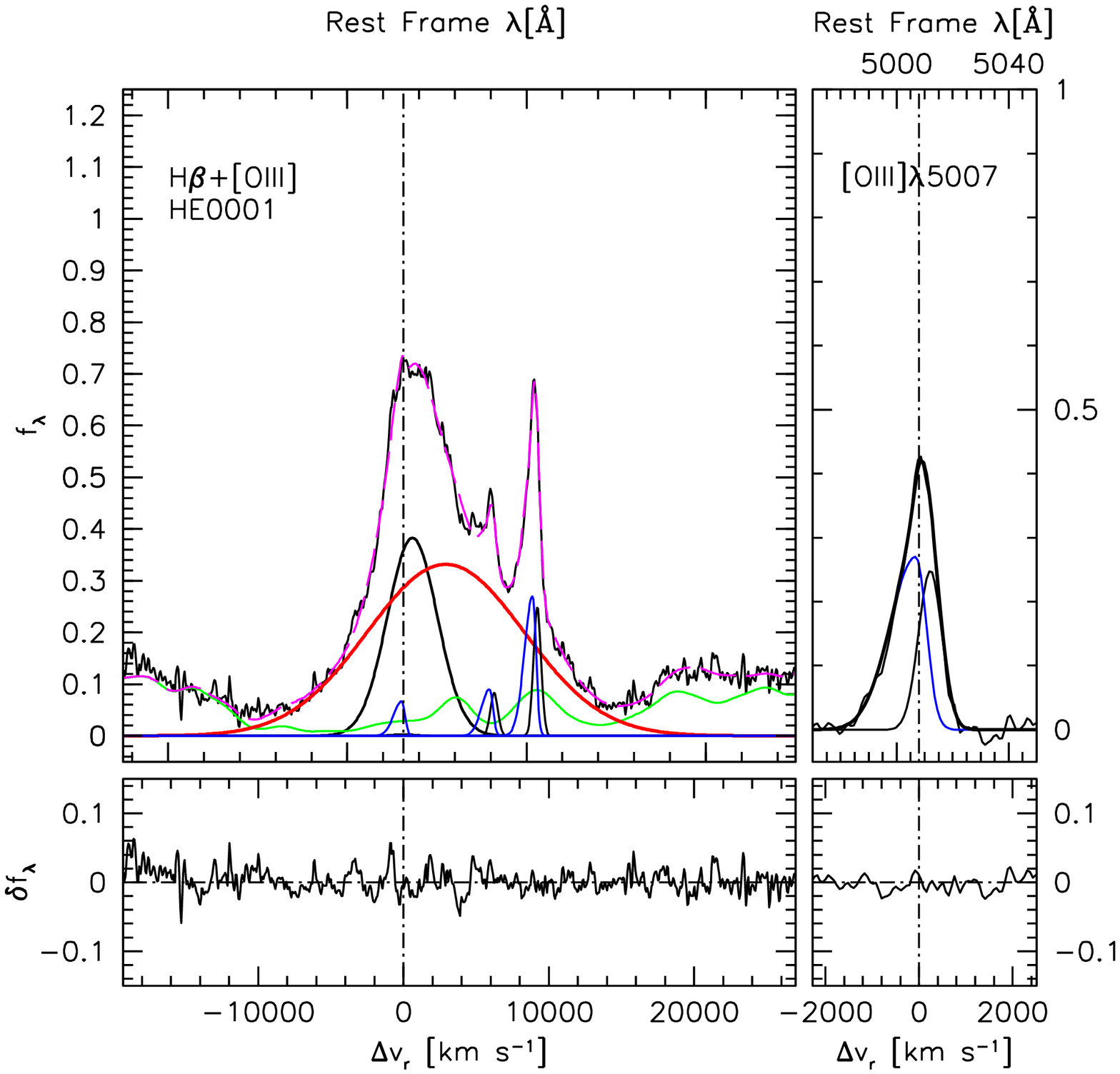}\\
\includegraphics[width=6.5 cm]{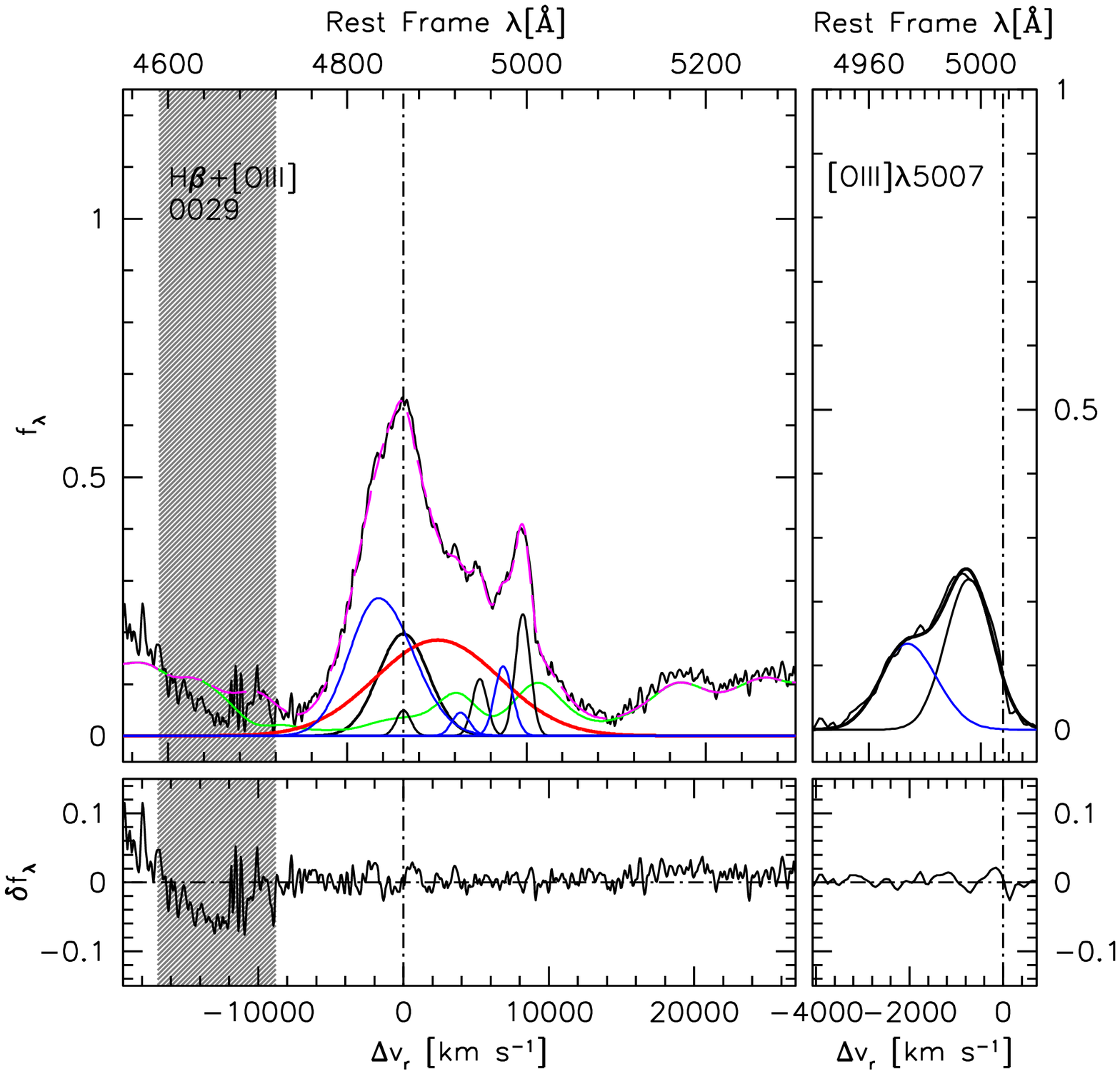}
 \caption{Analysis of the \hb\ + \oiiiopt\ region for two high-luminosity, high-$z$\ quasars belonging to the B1 spectral type. The top one, HE0001-234, shows no appreciable evidence of blueshift, while the bottom one [HB89] 0029+073 requires a stronger blue shifted excess for \oiiiopt, and an even stronger and broader one to fit \hb. Color coding of the components is the same as in the previous Figure. The shaded area identifies a spectral region affected by atmospheric absorptions.}
\label{fig:individuals} 
\end{figure}     
 
\begin{figure}[t!]
\centering
\includegraphics[width=6.5 cm]{b1m13.eps}\\
\includegraphics[width=6.5 cm]{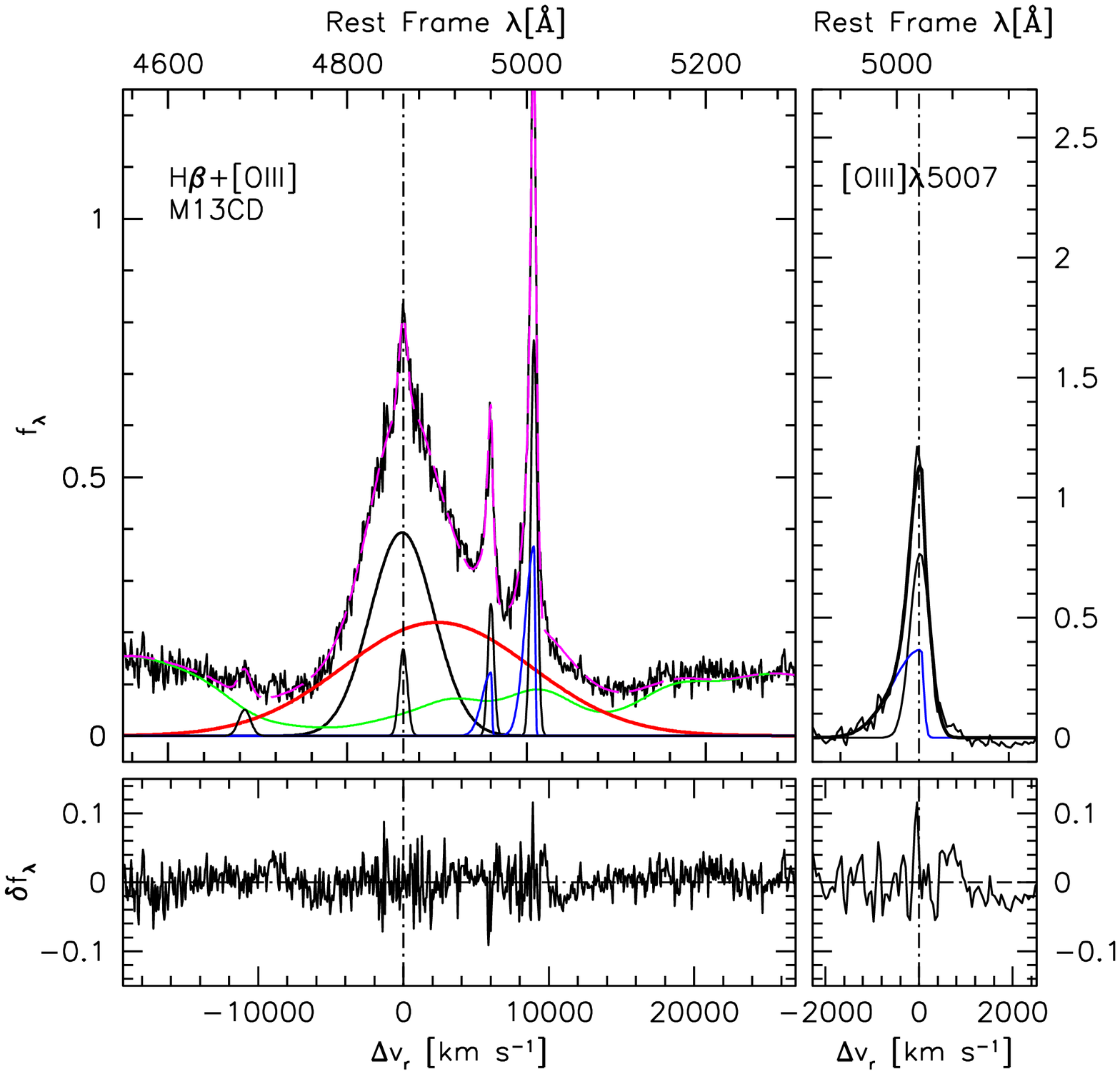}\\
\includegraphics[width=6.5 cm]{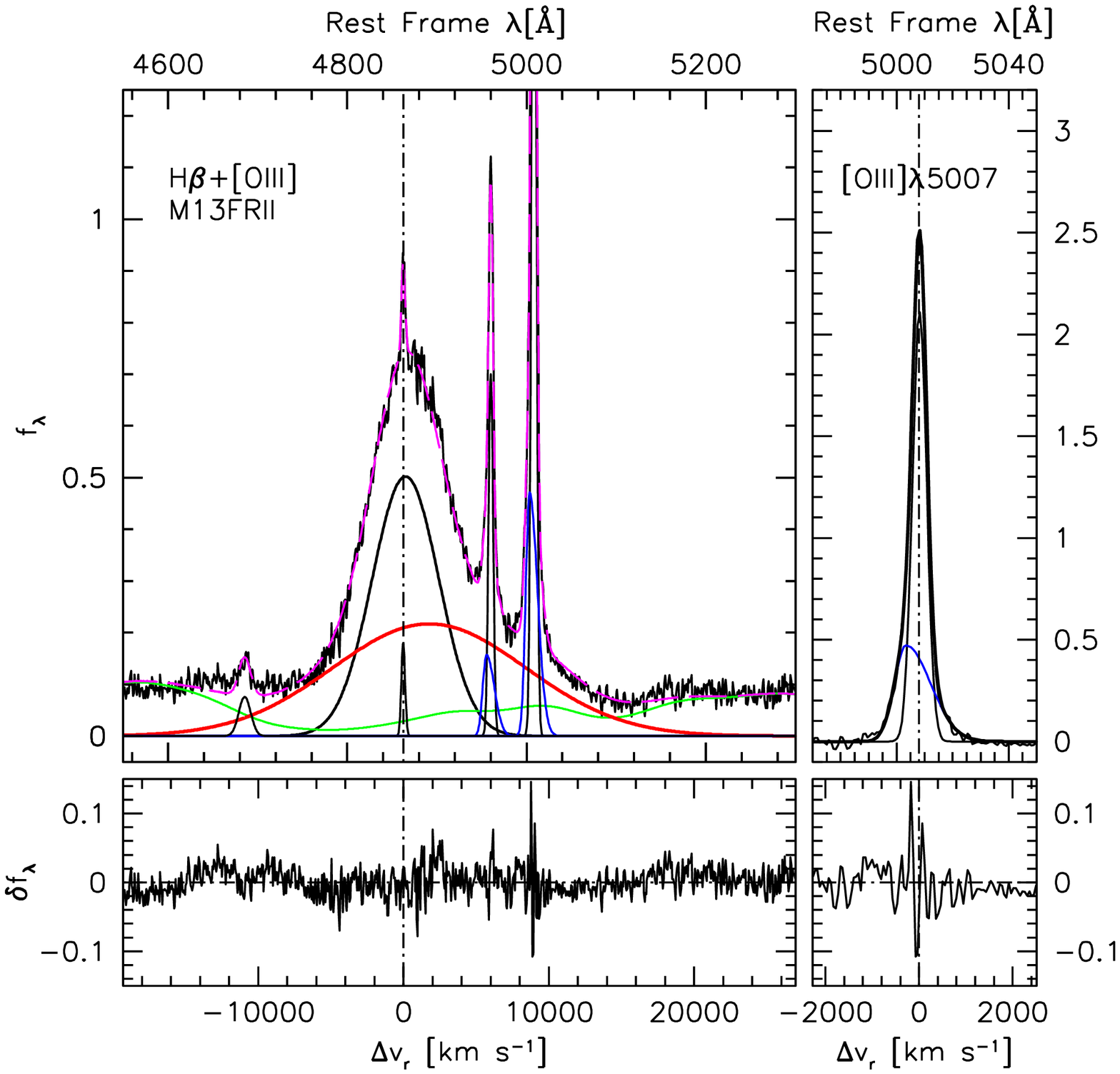}
 \caption{Analysis of the \hb\ + \oiiiopt\ region for the RQ composite spectrum of \citetalias{marzianietal13} (top), and for the CD and FR-II composite spectra (middle and bottom respectively). Color coding of the components is the same as in the previous Figures. }
\label{fig:rl} 
\end{figure}     

\paragraph{\oiii\ and \hb\ narrow-line emission} Table \ref{tab:mean} summarizes the measurements of the components associated with the narrow-line region (NLR) emission i.e.,  narrow and semi-broad components of \hb\ and \oiii\ (\hbnc, \oiiinc, and \hbsbc\ and \oiiisbc), for which normalized flux, equivalent width, shift and FWHM are reported. The skew parameter is reported only for the semi-broad components, as the narrow components are assumed to be symmetric Gaussian, within a few tens \kms\ from the  rest frame \citep{bonetal20}.  

%The last four Columns list the $F$ and $W$ of the \hbnc\ and of the \oiii. \textcolor{red}{[check the definitions]} Tables \ref{tab:hb} and Table \ref{tab:oiii} report the FWHM, the centroids and the $A.I.$\ as defined in \S \ref{fits}. 

The \citetalias{sulenticetal02} composite shows a broad + narrow component profile very well represented by three Gaussians: the symmetric unshifted \hbnc, the unshifted \hbbc\ and the \hbvbc\ with  a significant shift to the red. A small blue shifted excess appears at the interface between the \hbnc\ and the \hbbc, and has been modeled by an additional Gaussian. Its intensity is  so low that a very good fit with no significant worsening in the $\chi^{2}$\ can be achieved also without it. Most notably, the \oiii\ profile (enlarged in the right panel) is also fairly symmetric: a small centroid blueshift $\sim -50$ \kms is detected only at $\frac{1}{4}$\ peak intensity (Table \ref{tab:oiii}, where the \oiii\ full profile parameters are reported as in Table \ref{tab:hb} for \hb). The relatively large shift reported for \oiiisbc\ is compensated by a red-ward skew (Fig. \ref{fig:medians}, right panel on top row). In this case, the decomposition \oiiinc\ — \oiiisbc\ is especially uncertain, and a more reliable measurement is provided by the centroid. 

The \citetalias{marzianietal13} composite spectrum appears as  a ``goiter’’ at the top of the \hbbc\ broad profile. The \oiii\ profile is also fairly asymmetric, and can be modelled by a narrower, almost unshifted component and a skewed Gaussian displaced to the blue by $\approx -500$ \kms. The top of the \hb\ profile is well-fit by assuming two components with the same shift, width and asymmetry of the model components  \oiii\ line.  The consistency between the model of \hbsbc\ and \oiiisbc\ provide evidence that the \hb\ blueshifted and skewed component is associated with a NLR outflow.  The \citetalias{marzianietal09} composite can be equally modelled with the same skewed and blueshifted component for \hb\ and \oiii. However, this model would require an implausibly strong \oiiiopt\ emission. The fit shown in the bottom panel of Fig. \ref{fig:medians} assumes a broader component for the \hb\ emission.

 At very high luminosity (Fig. \ref{fig:individuals}) a prominent outflow is apparently absent in one Pop. B \hb\ profile (HE0001) but very prominent in another (Q0029). If the classification of Q0029 as a Population B source is correct, the model of the blue ``goiter’’ at the side of the \hb\ profile implies a strong contribution of blueshifted emission with a broad profile. The \oiii\ profiles are also different:  the equivalent width $W$ is higher and the shift lower in the case of HE0001, where no significant \hb\ outflow is detected. By all means, the  properties of Q0029 appear more extreme. We predict that this source will show extreme \civ\ blueshift, with amplitude of several thousands \kms.
 
The \oiii\ shift and the A.I. become more negative and the equivalent width decreases with increasing luminosity. This is a pure luminosity effect that goes in the same sense of the effect of increasing Eddington ratio in sample covering the full span of \lledd $\sim 10^{-2} - 1$, and can be interpreted as due to NLR evolution with redshift \citep{marzianietal16}. 
 
 %Note that again the extent of the \hb\ outflow seems to be related with the \oiii\ outflow, as for the cases shown in Fig. \ref{fig:medians}. 
 
 \paragraph{Jetted sources}  The CD and FR-II composites from the sample of \citetalias{marzianietal13}  (Fig. \ref{fig:rl}) show the \oiii\ blueshifted and skewed component is not detected in \hb, implying that for this component the intensity ratio \oiii/\hb $\gg1$.  In addition the \oiii\ profile for the FR-II composite spectrum is much more symmetric than that of the CD composite, whose $A.I.$\ and centroid shifts  are more consistent with the RQ composite of same sample. This systematic difference may arise because of the   different viewing angles expected for CD (seen almost pole on) and FR-II sources (seen at a viewing angle $\approx 40 - 60$\ \citep{urrypadovani95}).
 
\begin{table}
\begin{center}
\caption{Narrow line measurements \label{tab:mean}}\scriptsize\tabcolsep=1pt
 \hspace{-1cm}  \begin{tabular}{lccccccccccccccccccccc} \hline
Spectrum		&	\multicolumn{4}{c}{\hbnc}  & &  \multicolumn{5}{c}{\hbsbc}	& &	\multicolumn{4}{c}{\oiiinc} & &	\multicolumn{5}{c}{\oiiisbc} \\ \cline{2-5}\cline{7-11}\cline{13-16} \cline{18-22}
                         &      F     & W$^a$   & Shift$^b$ & FWHM$^b$ &&   F  &  W$^a$ & Shift$^b$ & FWHM$^b$ & Skew$^{c}$ &&     F     & W$^a$   & Shift$^b$ & FWHM$^b$ &&   F &  W$^a$ & Shift$^b$ & FWHM$^b$ & Skew$^{c}$  \\
%&               & [erg s$^{-1}$] &[AA] & \\ 
\hline
\\ \hline
\multicolumn{5}{c}{Composite spectra}\\
\hline  
B1S02	&	3.33	&	3.00	&	-9	&	492	&	&	0.59	&	0.53	&	-349	&	881	&	\ldots	&	&	14.6	&	14.0	&	12	&	499	&	&	8.86	&	8.45	&	-307	&	881	&	1.47	\\
B1M13	&	1.45	&	1.45	&	-8	&	450	&	&	0.83	&	0.85	&	-480	&	1054	&	0.26	&	&	8.9	&	9.7	&	11	&	513	&	&	2.76	&	3.00	&	-417	&	1054	&	0.26	\\
B1M09	&	0.65	&	0.70	&	-25	&	508	&	&	0.06	&	0.07	&	-752	&	932	&	1.46	&	&	1.6	&	1.8	&	-15	&	528	&	&	2.68	&	2.97	&	-688	&	932	&	1.46	\\  \hline
\multicolumn{5}{c}{Individual, high-$L$\ quasars}\\ 
\hline														  					
HE0001	&	0.11	&	0.10	&	-25	&	2202	&	&	1.05	&	0.99	&	-133	&	1301	&	0.41	&	&	2.6	&	2.7	&	239	&	592	&	&	4.39	&	4.43	&	-69	&	1301	&	0.41	\\
Q0029	&	1.00	&	0.95	&	-6	&	1181	&	&	0.00	&	0.00	&	\ldots	&	\ldots	&	\ldots	&	&	4.9	&	4.9	&	-728	&	1181	&	&	3.36	&	3.30	&	-2051	&	1340	&	1.12	\\  \hline
\multicolumn{5}{c}{Composite spectra, jetted}\\ \hline
B1M13CD	&	1.92	&	1.96	&	-25	&	662	&	&	0.00	&	0.00	&	\ldots	&	\ldots	&	\ldots	&	&	6.8	&	7.4	&	20	&	497	&	&	5.17	&	5.62	&	39	&	1439	&	0.10	\\
B1M13FRII	&	0.94	&	0.98	&	-25	&	300	&	&	0.00	&	0.00	&	\ldots	&	\ldots	&	\ldots	&	&	13.5	&	13.0	&	11	&	360	&	&	7.65	&	8.11	&	-279	&	585	&	2.12	\\ 
\hline
\end{tabular}
 \end{center}
\footnotesize{$^a$: in units of \AA; $^b$: in units of \kms.  $^{c}$ skew as reported by the  \textsc{specfit} routine; it is equal to the conventional definition of the skew \citep{azzaliniregoli12} + 1.}  
\end{table} 
\hoffset=0cm

\begin{table}
\begin{center}
\caption{\oiii\ profile measurements \label{tab:oiii}}\scriptsize\tabcolsep=5pt
 \hspace{-1cm}  
 \begin{tabular}{lcccccc} \hline
Spectrum	&	FWHM$^{a}$		&	AI	&	c(1/4)$^{a}$	&	c(1/2)$^{a}$	&	c(3/4)$^{a}$		&	c(0.9)$^{a}$	 	\\
\hline
\multicolumn{7}{c}{Composite spectra}\\
\hline  
B1S02	&	580	$\pm$	30	&	-0.10	$\pm$	0.08	&	-40	$\pm$	40	&	-10	$\pm$	20	&	0	$\pm$	20	&	0	$\pm$	10	\\
B1M13	&	560	$\pm$	40	&	-0.23	$\pm$	0.11	&	-80	$\pm$	50	&	-20	$\pm$	20	&	10	$\pm$	10	&	30	$\pm$	10	\\
B1M09	&	1100	$\pm$	120	&	-0.43	$\pm$	0.05	&	-380	$\pm$	40	&	-280	$\pm$	60	&	-60	$\pm$	20	&	-40	$\pm$	10	 	\\  \hline
\multicolumn{7}{c}{Individual, high-$L$\ quasars}\\ 
\hline	
HE0001	&	900	$\pm$	70	&	-0.26	$\pm$	0.07	&	-100	$\pm$	40	&	0	$\pm$	30	&	70	$\pm$	30	&	80	$\pm$	10	\\
Q0029	&	2120	$\pm$	140	&	-0.37	$\pm$	0.04	&	-1360	$\pm$	60	&	-1240	$\pm$	70	&	-860	$\pm$	50	&	-830	$\pm$	30	 	\\ \hline
\multicolumn{7}{c}{Composite spectra, jetted}\\ \hline
B1M13CD	&	490	$\pm$	40	&	-0.18	$\pm$	0.12	&	-90	$\pm$	50	&	-70	$\pm$	20	&	-10	$\pm$	20	&	-10	$\pm$	10	\\
B1M13FRII	&	440	$\pm$	30	&	-0.10	$\pm$	0.09	&	-20	$\pm$	30	&	10	$\pm$	10	&	10	$\pm$	10	&	10	$\pm$	10	\\
\hline
\end{tabular}
  \end{center}
\footnotesize{\footnotesize{$^a$: in units of \kms. }  }  
\end{table} 
\hoffset=0cm	

\section{Discussion}
\label{discussion}
 
The analysis performed above has been focused on sources radiating at relatively modest \lledd\ (Population B) but covering a wide range of redshifts (0 $\lesssim z \lesssim 3$) and luminosities. Significant outflow features have been detected in the NLR, as traced by the \hbnc\ and \oiiiopt\ blue shifted components. At high luminosity, significant blueshift are found not only in the \oiiiopt\ lines, but also with a broader profile, hinting at an association with the BLR emission.  

\subsection{How important is the outflow component?}

The present analysis relies on the important assumption that the Population B profile at \hb\ low-$z$ and luminosity is not significantly affected by  any outflowing gas. Reverberation mapping campaigns in the early 2000s provided evidence that the main broadening mechanism is indeed provided by a virial velocity field of gas orbiting around a point-like mass. More recent works points toward a more complex situation \citep[][Bao et al. 2022, in preparation]{denneyetal09a,duetal18a,uetal22}, although the main inference from velocity-resolved reverberation mapping studies for the sources with the red \hb\ asymmetry is that the velocity field is predominantly virial, with the frequent detection of infall motions.  The detection of infall is based on the shorter time delay of the red wing, not on the response of the line core. 

\subsection{Identifying an outflow component}

The \hb\ profile of Population B presents a clear inflection between \hbbc\ and \hbnc\ that can be explained on the basis of the expected radial emissivity of \hb\ \citep{sulenticmarziani99}.   The identification of an outflow component may be achieved by considering the following options:
\begin{itemize}
\item no significant centroid blueshift in the broad profile of \hb\ and symmetric  appearance at the interface between \hbnc\ and \hbbc, with the peak of the broad profile showing no shift or a slight redshift: no evidence of outflow. 
\item No significant centroid blueshift in the broad profile of \hb\ and ``goiter’’ appearance at the interface between \hbnc\ and \hbbc: If the \oiii\ line shows a significant blueward asymmetry, and a model of the \oiii\ line profile with a core and semi-broad component is applicable to the \hb\ profile, then it is likely that the outflow is mainly associated with the NLR emission.
\item Even modest centroid blueshift in the broad profile of \hb\ at fractional intensity  $\frac{3}{4}$\  or 0.9, the outflow might involve BLR emission. In this case, the \hbblue\ corresponds to the prominent blueshifted  emission of the \civ\ line observed at high luminosity \citep{sulenticetal17}. The detection of \hbblue\ is made more difficult by the \civ/\hb\ ratio expected to be  $\gg 1$.
\end{itemize}

\subsection{Location and physical nature of the outflow}
 
Even in case of modest accretion rate, the outflow can be radiatively driven \citep{netzermarziani10}. The ratio between the radiation and gravitation force can be written as $a_\mathrm{rad} / a_\mathrm{grav} \approx 7.2 $ \lledd\ $N_\mathrm{c,23}^{-1}$\ where  $N_\mathrm{c,23}$\ is the Hydrogen column density in units of $10^{23}$ cm$^{-2}$ \ \citep[e.g.,][]{ferlandetal09}.  For \lledd\ $\sim 0.1$, gas of moderate common density $N_\mathrm{c,23} \sim 0.1 $ could be accelerated to  $a_\mathrm{rad} / a_\mathrm{grav} \sim 10$ \citep[c.f. Eq. 6 of \citet{netzermarziani10}][]{marzianietal10}. The first underlying assumption is that all of the photon momentum of the ionizing continuum is transferred to the line emitting gas. The second assumption is that the gas is optically thick to the ionizing continuum, and this condition is more easily verified if the ionization parameter is low, implying that the low column density gas located farther out from the AGN continuum source might be preferentially accelerated.  This might explain why we see a signature due to a semi-broad component  in \hb, \hbsbc, in turn associated with the \oiii\ semi-broad component, likely at the inner edge of the NLR, may be the main signature of outflow in low \lledd\ sources. 

Regarding the BLR, at low luminosity there is no signature of outflow, if our interpretation of the profile is correct. For Population B sources, however, the observed spectrum can be explained by the locally optimized cloud (LOC) scheme, in which a range of ionization parameters, density and column density is assumed, and the emerging spectrum is set by the parameters at which lines are emitted most efficiently \citep{baldwinetal95,koristaetal97}. This is to say that there might be always gas as ``light’’ as needed for an outflow; however, that outflow may not produce a significant signature in the emission line spectrum.  Powerful outflow at modest Eddington ratio may become possible only at high luminosity \citep[e.g.,][]{murraychiang97,progaetal98,laorbrandt02}, as predicted from wind theory, and confirmed by observations  \citep{bischettietal17,vietrietal17,vietrietal18,sulenticetal17}.

\subsection{The fate of the outflowing gas: no feedback effects at low $L$}

The {mass outflow rate} at a distance $r$\  can be written as, if the flow is confined to a solid angle of $\Omega$ of volume $\frac{4}{3}\pi r^{3} \frac{\Omega}{4 \pi} $: $ \dot{M}^\mathrm{ion}_\mathrm{o} =  \rho\ \Omega r^{2} v_\mathrm{o}  = \frac{{M}^\mathrm{ion}_\mathrm{o}}{V} 	\Omega r^{2} v_\mathrm{o}    \propto  L  v_\mathrm{o}  r^{-1} $ \citep{canodiazetal12}, and implies 
$\dot{M}^\mathrm{ion} \sim 30  L_{44} v_\mathrm{o,1000} r^{-1}_{\rm 1 kpc}   \left(\frac{Z}{5Z_{\odot}}\right)^{-1} n_{3}^{-1}$, where the mass of ionized gas can be directly estimated from the line luminosity:  $M_\mathrm{ion} \sim 1 \cdot 10^7~L_{44} \left(\frac{Z}{5Z_{\odot}}\right)^{-1} n_{3}^{-1} $. \footnote{Note that the filling factor is not appearing explicitely because by using the line luminosity we already are considering the volume of the line emitting gas. The fraction of volume that is actually occupied by the line emitting gas then depends on its density.}   The low-luminosity cases \citetalias{sulenticetal02} and \citetalias{marzianietal13} imply that the outflow velocity is $v_\mathrm{o,1000} \sim 1$\ from the peak shift of \oiiisbc, and the \oiiisbc\ luminosity is $\log L_\mathrm{[OIII]} \sim 42$.  Assuming $Z \approx 1 Z_{\odot}$\ as appropriate for Population B sources \citep{punslyetal20},  $M_\mathrm{ion} \sim 5 \cdot 10^5   n_{3}^{-1} $, and $\dot{M}^\mathrm{ion} \sim 0.15  r^{-1}_{\rm 1 kpc}  n_{3}^{-1}$. By the same token  the thrust and kinetic power can be written as $ \dot{M}v \sim 1.9 \cdot 10^{35} L_{44}  v^{2}_\mathrm{o,1000} r^{-1}_{\rm 1 kpc}    \left(\frac{Z}{5Z_{\odot}}\right)^{-1} n_{3}^{-1} $ and   $\dot{\epsilon} \sim 10^{43} L_{44}  v^{3}_\mathrm{o,1000}   r^{-1}_{\rm 1kpc}  \left(\frac{Z}{5Z_{\odot}}\right)^{-1} n_{3}^{-1}   $, which become $ \dot{M}v \sim 1 \cdot 10^{34}  r^{-1}_{\rm 1 kpc}   n_{3}^{-1} $\ and $\dot{\epsilon} \sim 5 \cdot 10^{41}  r^{-1}_{\rm 1kpc}  n_{3}^{-1}$. Even assuming that we are observing a flow at  $r \sim 10$ pc, the  kinetic power is $\dot{\epsilon} \lesssim 10^{44}$ \ergss, a factor $\approx 100$ below the bolometric luminosity and  $\approx 1000$\ the Eddington luminosity of the \citetalias{marzianietal13} case. The emitting gas might be beyond or at the limit of the black hole sphere of influence given by $r \approx GM /\sigma_{\star}^{2} \approx 8 \cdot 10^{19} M_{9,\odot} /\sigma_{\star,400}^{2}$ cm, where the $\sigma_{\star}$ is the velocity dispersion associated with the bulge of the host galaxy in units of 400 \kms. At this radius, the escape velocity is expected to be $v_\mathrm{esc} \sim 500 $ \kms \ for a $10^{9}$ \msol\ black hole. It is therefore doubtful whether the outflowing gas might be even able to escape  from the sphere of influence of the black hole. Even less likely is that the outflowing gas might ``wreak havoc’’ galaxy-wide in the bulge and disk of the host, due to the small amount of gas masses involved in the outflow, and due to the escape velocity that can be as high as $v_\mathrm{esc} \gtrsim 1000$ \kms\ in the inner regions of a massive spheroid or in a giant spiral such as the Milky Way \citep{monarietal18}. 

The scenario might radically change at high luminosity: considering the \citetalias{marzianietal09} composite, the velocity of \oiiisbc\ is higher by a factor $\approx 2$, and the line luminosity by a factor $\sim 10$, implying a 20-, 40, $\sim$ 100-fold increase  over the \citetalias{marzianietal13} case in mass flow, thrust and kinetic power, respectively. In the \citetalias{marzianietal09} case the kinetic power would be comparable to the Eddington luminosity. An even more powerful outflow is expected for Q0029.  

\section{Summary and conclusion}

The analysis of outflow signatures carried out in the present paper has been focused on three samples of type-1 AGN covering a wide range of luminosity.  

The detection of different kinematic components in single epoch profiles is a complicated issue. The apertures and slit widths used in ground-based observation add up the emission from AGN continuum, BLR, NLR, and host galaxy that are associated with widely different spatial scales. The case of Population B sources of spectral type B1 is especially well-suited to analyze the presence of an outflow component in the Balmer \hb\ line for sources that are radiating at modest Eddington ratio.  

Generally speaking, the detection of significant systematic blueshifts in the centroid measurements can be taken as a signature of outflow. If the blueshift/blue asymmetry is confined at the top of the \hb\ line, and the \hb\ narrow emission can be modeled as \oiii\ assuming a semi-broad and a narrow component with a similar parameter, then the evidence of the outflow  (the ``goiter’’ in the line profile) remains confined to the NLR. However, if the \hb\ centroid at $\frac{3}{4}$\ or at lower fractional intensity is also blue shifted, it is likely that a BLR outflow is being detected. Low column density gas can be driven into an outflow by radiation forces. Blueshifts in the line core can be therefore straightforwardly interpreted by an outflow component, without invoking binary BLR, in turn pointing toward sub-parsec binary black holes. Other spectral types along the MS have been identified as frequently involving binary black hole candidates \citep{gancietal19,delolmoetal21}. 

The estimates of mass flow, thrust and kinetic power are highly uncertain because of the lack of spatially-resolved data. This situation might be changing soon with the development of integral-field spectrographs. Nonetheless, even maximizing the coarse estimates reported above, it is unlikely that the thrust and the kinetic power (just $\sim 10^{-2}$\ the Eddington luminosity as derived for the \citetalias{sulenticetal02} and \citetalias{marzianietal13} samples) might have a strong impact on the host galaxy evolution, not to mention the possibility of driving the black hole mass — bulge correlation \citep[e.g.,][and references therein]{donofrioetal21}. Even if the \oiii\ samples only emission from mildly ionized gas, and the mass flow  might be  dominated by the higher-ionization gas,  for low luminosity AGN such as the prototypical Population B Seyfert-1 NGC 5548 the kinetic luminosity remains a very small fraction of the Eddington luminosity \citep{kaastraetal14,kriss17}.  The situation is expected to change at the ``cosmic noon’’ at redshifts in the range 1 — 2, when the most luminous quasars are observed, and of which  \citetalias{marzianietal09} provides a representative spectrum.

%In most cases, the outflowing gas is located at parsec to kpc scales, with insufficient kinetic luminosity to have an evolutionary impact on the host galaxy. Typically, the kinetic luminosity is less than a percent of the Eddington luminosity. 

\authorcontributions{{All authors contributed equally to this paper.}}

%%%%%%%%%%%%%%%%%%%%%%%%%%%%%%%%%%%%%%%%%%
%\funding{{P.M.} wishes to thank Ascensi{o}n del Olmo and Jaime Perea for the allotted time on their departmental server {\tt hypercat} at IAA-CSIC.  }

%%%%%%%%%%%%%%%%%%%%%%%%%%%%%%%%%%%%%%%%%%
\acknowledgments{A.D.M. and A.d.O. acknowledge financial support from the State Agency for Research of the Spanish MCIU through the project PID2019-106027GB-C41 and the “Center of Excellence Severo Ochoa” award to the Instituto de Astrofísica de Andalucía (SEV-2017-0709). A.D.M. acknowledges the support of the INPhINIT fellowship from "la Caixa" Foundation (ID 100010434). The fellowship code is LCF/BQ/DI19/11730018.

Funding for the Sloan Digital Sky Survey has been provided by the Alfred P. Sloan Foundation, and the U.S. Department of Energy Office of Science. The SDSS web site is \texttt{http://www.sdss.org}. SDSS-III is managed by the Astrophysical Research Consortium for the Participating Institutions of the SDSS-III Collaboration including the University of Arizona, the Brazilian Participation Group, Brookhaven National Laboratory, Carnegie Mellon University, University of Florida, the French Participation Group, the German Participation Group, Harvard University, the Instituto de Astrofisica de Canarias, the Michigan State/Notre Dame/JINA Participation Group, Johns Hopkins University, Lawrence Berkeley National Laboratory, Max Planck Institute for Astrophysics, Max Planck Institute for Extraterrestrial Physics, New Mexico State University, University of Portsmouth, Princeton University, the Spanish Participation Group, University of Tokyo, University of Utah, Vanderbilt University, University of Virginia, University of Washington, and Yale University.}
%%%%%%%%%%%%%%%%%%%%%%%%%%%%%%%%%%%%%%%%%%
\conflictsofinterest{The authors declare no conflict of interest.} 

%%%%%%%%%%%%%%%%%%%%%%%%%%%%%%%%%%%%%%%%%%
%% optional
\abbreviations{The following abbreviations are used in this manuscript:\\
\noindent 
\begin{tabular}{@{}ll}

AGN& Active Galactic Nucleus/i\\
BLR& Broad Line Region\\
BC & Broad Component\\
{CD}& {Core Dominated} \\
%{FR-I}& {Fanaroff-Riley I} \\ 
{FR-II}& {Fanaroff-Riley II} \\
FWHM & Full Width Half-Maximum\\
{HE} & {Hamburg-ESO} \\
{ISAAC} &{ Infrared Spectrometer And Array Camera}\\
IR & {Infrared}\\
{LOC}& {Locally Optimized Cloud}\\
%FIZ & Fully Ionized Zone\\
%DESI& Dark Energy Spectroscopic Instrument \\
%ESC& Eddington Standard Candles\\
%HIL& High-Ionization Line\\
%LIL& Low-Ionization Line\\
M03 & \citeauthor{marzianietal03a}\citeyear{marzianietal03a}\\
M09 & \citeauthor{marzianietal09}\citeyear{marzianietal09}\\
M13 & \citeauthor{marzianietal13a}\citeyear{marzianietal13a}\\
MDPI& Multidisciplinary Digital Publishing Institute\\
MS& Main Sequence\\
NC& Narrow Component\\
NGC & New General Catalogue\\
NLR& Narrow Line Region\\
%NLSy1 & Narrow-Line Seyfert 1\\
%PIZ & Partially Ionized Zone \\
{RL} & {Radio loud} \\
{RQ} & {Radio quiet} \\
SBC& Semi-Broad Component\\
{SDSS} & {Sloan Digital Sky Survey}\\
%{SED} & {Spectral energy distribution}\\
%{S/N} & {Signal-to-noise ratio}\\
UV & Ultra-violet \\
VBC& Very Broad Component\\
VBLR & Very Broad Line Region\\
\end{tabular} }

\vfill\eject\pagebreak\newpage

%=====================================
% References, variant B: external bibliography
%=====================================

%\externalbibliography{yes}
%\begin{thebibliography}{999}
%\providecommand{\natexlab}[1]{#1}
\bibliography{biblioletter2}
%\begin{thebibliography}{-------}
%\providecommand{\natexlab}[1]{#1}

%\end{thebibliography}
%
%%%%%%%%%%%%%%%%%%%%%%%%%%%%%%%%%%%%%%%%%%
%% optional
%\sampleavailability{Samples of the magic compounds needed to turn quasars into standard candles are {\em not} available from the authors.}

%% for journal Sci
%\reviewreports{\\
%Reviewer 1 comments and authors’ response\\
%Reviewer 2 comments and authors’ response\\
%Reviewer 3 comments and authors’ response
%}
%%%%%%%%%%%%%%%%%%%%%%%%%%%%%%%%%%%%%%%%%%
\end{document}